\documentclass[12pt]{article} 
\usepackage[a4paper,textwidth=490.0pt,textheight=703.1pt]{geometry}
\usepackage{url}
\usepackage{booktabs}

\usepackage{epsfig} %
\usepackage{float} %
\usepackage{amssymb} %
\usepackage{amsmath} %
\usepackage{verbatim} %
\usepackage{cite} %
\usepackage{color} %
\usepackage{eufrak}
\usepackage[english]{babel}
\usepackage[latin1]{inputenc}
\usepackage[T1]{fontenc}
\usepackage{ae}                    
\usepackage{url}   
\usepackage{graphicx} 
\usepackage{booktabs}
\usepackage{slashed}

\newcommand{\HA}{{\rm H}}

\newcommand{\ep}{\varepsilon}

\newcommand{\beq}{\begin{equation}}
\newcommand{\eeq}{\end{equation}}
\newcommand{\bea}{\begin{eqnarray}}
\newcommand{\eea}{\end{eqnarray}}

\allowdisplaybreaks[4]

\begin{document} 
\setlength{\baselineskip}{0.515cm}

\sloppy 
\thispagestyle{empty} 
\begin{flushleft} 
DESY 21--028
\\ 
DO--TH 21/33
\\ 
TTP 21--007
\\ 
RISC Report Series 22--01
\\ 
SAGEX--21--39
\\ 
February 2022
\end{flushleft}

\mbox{} \vspace*{\fill} \begin{center}

{\LARGE\bf The Two-Loop Massless Off-Shell QCD}

\vspace*{2mm} 
{\LARGE\bf Operator Matrix Elements to Finite Terms}

\vspace{3cm} 
\large 
{\large J.~Bl\"umlein$^a$, P.~Marquard$^a$, C.~Schneider$^b$ and K.~Sch\"onwald$^{c}$ }

\normalsize 

\vspace{1.cm} 
{\it $^a$~Deutsches Elektronen--Synchrotron DESY,}\\ {\it Platanenallee 6, 15738 Zeuthen, Germany}

\vspace*{2mm} 
{\it $^b$~Research Institute for Symbolic Computation (RISC), Johannes Kepler Universty, Altenbergerstra\ss{}e 69, A-4040 Linz, 
Austria}

\vspace*{2mm} 
{\it $^c$~Institut f\"ur Theoretische Teilchenphysik,\\ Karlsruher Institut f\"ur Technologie (KIT) D-76128 
Karlsruhe, Germany}


\end{center} 
\normalsize 
\vspace{\fill} 
\begin{abstract} 
\noindent
We calculate the unpolarized and polarized two--loop massless off--shell operator matrix elements in QCD 
to $O(\ep)$ in the dimensional parameter in an automated way. Here we use the method of arbitrary high 
Mellin moments and difference ring theory, based on integration-by-parts relations. This method also 
constitutes one way to compute the QCD anomalous dimensions. The presented higher order contributions to 
these operator matrix elements occur as building blocks in the corresponding higher order calculations up
to four--loop order. All contributing quantities can be expressed in terms of harmonic sums in Mellin--$N$ 
space or by harmonic polylogarithms in $z$--space. We also perform comparisons to the literature. 
\end{abstract}

\vspace*{\fill} \noindent
\newpage

\section{Introduction} 
\label{sec:1}

\vspace*{1mm} 
\noindent 
The unpolarized and polarized anomalous dimensions of the local twist--2 operators in Quantum Chromodynamics (QCD) 
play a fundamental role in the description of the scaling violations of the deep--inelastic structure functions.
Their measurement provides one of the safest ways to measure the strong coupling constant $\alpha_s(M_Z^2) = 4 \pi 
a_s$ \cite{alphas}. While at first order, there is a wide variety of possibilities to calculate the anomalous 
dimensions, see e.g.~\cite{REV}, at higher orders only a few efficient methods are known. These are based either 
on the calculation of massless off--shell operator matrix elements (OMEs) \cite{Gross:1973ju,Gross:1974cs,
Georgi:1951sr,Sasaki:1975hk,Ahmed:1975tj,Floratos:1977au,GonzalezArroyo:1979he,GonzalezArroyo:1979ng,
GonzalezArroyo:1979df,Curci:1980uw,Furmanski:1980cm,Floratos:1981hs,Hamberg:1991qt,Mertig:1995ny,SP_PS1,
Ellis:1996nn,Matiounine:1998ky,Matiounine:1998re}, the calculation of the forward Compton amplitude 
of a space--like virtual gauge boson on a massless on--shell parton 
\cite{Moch:1999eb,Moch:2004pa,Vogt:2004mw,
Vogt:2008yw,Moch:2014sna} and on massive on--shell OMEs \cite{Buza:1995ie,Buza:1996xr,Buza:1996wv,Bierenbaum:2007qe,
Bierenbaum:2009zt,POL19,PVFNS,Ablinger:2010ty,Ablinger:2014vwa,Ablinger:2014nga,Ablinger:2014lka,Behring:2014eya,
Ablinger:2017tan,Ablinger:2019etw,Behring:2019tus,LOGPOL}. All these methods have advantages and disadvantages and 
form complimentary ways to compute the anomalous dimensions.

To perform phenomenological analyses of the deep--inelastic world data also the process dependent massless and 
massive Wilson coefficients have to be calculated \cite{Furmanski:1981cw,Kodaira:1978sh,Kodaira:1979pa,
UNPOL,Alekhin:2003ev,Kazakov:1987jk,Bodwin:1989nz,Kazakov:1990fu,SanchezGuillen:1990iq,vanNeerven:1991nn,
Zijlstra:1991qc,Zijlstra:1992qd,Zijlstra:1992kj,NUMTL,Zijlstra:1993sh,Moch:1999eb,Vermaseren:2005qc,Moch:2008fj,
HWILS1,Buza:1995ie,Gluck:1996ve,Buza:1997mg,Blumlein:2011zu,Behring:2014eya,Blumlein:2014fqa,Behring:2015roa,
Behring:2015zaa,Blumlein:2016xcy,Behring:2016hpa,Hekhorn:2018ywm,Blumlein:2019qze,Blumlein:2019zux,POL19,LOGPOL}.
This is also necessary to account for the specific scaling violations implied by massive quarks, such 
as charm and bottom.

In this paper we present the results for the unpolarized and polarized massless off-shell OMEs up to two--loop 
order and to corrections of $O(\ep)$ (at $O(a_s)$ of $O(\ep^2)$), with the dimensional parameter $\ep = D-4$,
including non--gauge invariant contributions. A part of
these expansion coefficients of the various OMEs contribute to the physical OMEs up to four--loop order. 
Others emerge due to the kinematic breaking of gauge invariance both in the unpolarized and polarized 
off--shell case. One characteristics
is the emergence of additional OMEs related to the breaking of the equation of motion (eom) and of new 
non--gauge invariant OMEs, also with new unphysical anomalous dimensions. These quantities play a role in the
calculation of the unpolarized anomalous dimensions due to mixing. We extend earlier work of 
Refs.~\cite{Matiounine:1998ky,Matiounine:1998re} and perform the calculation in an automated way, applying 
methods which have been developed by us solving a series of massive three--loop problems and other applications 
during the last decade.

The theoretical basis for these calculations has been laid out in a series of papers describing the situation 
holding in the off--shell case, breaking gauge invariance, cf.~\cite{Gross:1974cs,Dixon:1974ss,
KlubergStern:1974xv,Sarkar:1974db,Sarkar:1974ni,Joglekar:1975nu,Joglekar:1976eb,Hamberg:1991qt,
Collins:1994ee,Harris:1994tp,Matiounine:1998ky,Matiounine:1998re}. Compared to the method of the forward 
Compton amplitude, the method of massless off--shell OMEs needs no precautions as reference to Higgs and 
gravitational subsidiary fields. The use of massive on--shell OMEs, as also the method of the forward Compton 
amplitude do not encounter gauge invariance problems, on the other hand. However, the massive OMEs allow to 
derive only the contributions $\propto T_F$ of the anomalous dimensions, except of going to one order higher 
in the coupling constant. A new challenge in the present approach is to master the breaking of gauge invariance 
to the respective perturbative order. We present the results in Mellin--$N$ space, because the expressions 
are 
more compact than in momentum fraction $z$ space. We also perform a detailed comparison to the literature
\cite{Matiounine:1998ky,Matiounine:1998re,Hamberg:1991qt,HAMBERG} and correct results given there, including 
Feynman rules, and also all non--gauge invariant terms. A by--product of the present calculation is the 
calculation of the (by now well--known) unpolarized physical anomalous dimensions to two--loop order. We present 
all contributing expansion coefficients up to $O(a_s^2)$ to the depth being needed in future four--loop 
calculations, extending
the level previously attempted in Refs.~\cite{Matiounine:1998ky,Matiounine:1998re}.

The paper is organized as follows. In Section~\ref{sec:2} we give a brief outline of the formalism, including 
the renormalization, and the main steps of the calculation of the different off--shell OMEs. We then turn 
to the calculation of the OMEs of the so-called alien operators in Section~\ref{sec:3}, which also lead to
additional anomalous dimensions. These operator matrix elements contribute via mixing to the 
renormalization of the unpolarized singlet operators, which we discuss in Section~\ref{sec:4}.
In Sections~\ref{sec:5}--\ref{sec:7} we present the expansion coefficients
of the unpolarized and the polarized standard OMEs and those for transversity for non--negative powers in the 
dimensional parameter $\ep$. In Section~\ref{sec:8} we compare to and correct partial results given 
previously 
in Refs.~\cite{Matiounine:1998ky,Matiounine:1998re}. Section~\ref{sec:9} contains the conclusions.
In Appendix~\ref{sec:A} we provide Feynman rules for the alien operators, if not previously being presented in
Refs.~\cite{Bierenbaum:2009mv,Behring:2019tus,Blumlein:2021ryt}, and list the contributing 
polynomials appearing in the standard OMEs in Appendix~\ref{sec:B}. Our results both in Mellin--$N$ space 
and 
$z$--space, are given in ancillary files in computer--readable form. In $z$--space we use the decomposition 
given in \cite{Blumlein:2021enk}, Eqs.~(45,46).
\section{The Formalism} 
\label{sec:2}

\vspace*{1mm} 
\noindent 
The massless off--shell OMEs are defined by
\begin{eqnarray}
\hat{A}_{ij}^{l} = \langle j(p)|O_i^l|j(p)\rangle,~~~~~\text {with}~~~i,j = q,g,
\end{eqnarray}
where $l$ further labels the type of the OME and $p$ denotes the off--shell momentum with $p^2 < 0$. 
The twist--2 local physical operators\footnote{The unphysical local operators are defined in 
Section~\ref{sec:3} below.} are given by
\begin{eqnarray}
O_{q,r; \mu_1 ... \mu_N}^{\sf NS} &=&
i^{N-1} {\rm\bf S}\Biggl[
\overline{\psi} \gamma_{\mu_1} D_{\mu_2} ... D_{\mu_N} \frac{\lambda_r}{2} \psi \Biggr] 
-~\text{trace~terms},
\label{eq:OP1}
\\
O_{q; \mu_1 ... \mu_N}^{\sf S} &=&
i^{N-1} {\rm\bf S}\Biggl[
\overline{\psi} \gamma_{\mu_1} D_{\mu_2} ... D_{\mu_N}  \psi \Biggr] -~\text{trace~terms},
\label{eq:OP2}
\\
O_{g; \mu_1 ... \mu_N}^{\sf S} &=&
2 i^{N-2} 
{\rm\bf S}
{\rm\bf Sp}
\Biggl[F_{\mu_1 \alpha}^a D_{\mu_2} ... D_{\mu_{N-1}} F_{\mu_N}^{\alpha,a} \Biggr] -~\text{trace~terms},
\label{eq:OP3}
\end{eqnarray}
in the unpolarized case. In the polarized case the operators are
\begin{eqnarray}
O_{q,r; \mu_1 ... \mu_N}^{\sf NS} &=&
i^{N-1} {\rm\bf S}\Biggl[
\overline{\psi} \gamma_5  \gamma_{\mu_1} D_{\mu_2} ... D_{\mu_N} \frac{\lambda_r}{2} \psi 
\Biggr] -~\text{trace~terms},
\label{eq:OP1}
\\
O_{q; \mu_1 ... \mu_N}^{\sf S} &=&
i^{N-1} {\rm\bf S}\Biggl[
\overline{\psi} \gamma_5 \gamma_{\mu_1} D_{\mu_2} ... D_{\mu_N}  \psi \Biggr] -~\text{trace~terms},
\label{eq:OP2p}
\\
O_{g; \mu_1 ... \mu_N}^{\sf S} &=&
2 i^{N-2} 
{\rm\bf S}
{\rm\bf Sp}
\Biggl[\frac{1}{2} \ep_{\mu_1 \alpha \beta \gamma}
F^{\beta \gamma,a} D_{\mu_2} ... D_{\mu_{N-1}} F_{\mu_N}^{\alpha,a} \Biggr] -~\text{trace~terms}.
\label{eq:OP3p}
\end{eqnarray}
For transversity \cite{Blumlein:2009rg} the following local non--singlet operator contributes
\begin{eqnarray}
O_{q,r; \mu \mu_1 ... \mu_N}^{\sf NS} &=&
i^{N-1} {\rm\bf S}\Biggl[
\overline{\psi} \sigma_{\mu \mu_1} D_{\mu_2} ... D_{\mu_N} \frac{\lambda_r}{2} \psi \Biggr] 
-~\text{trace~terms},
\label{eq:TR}
\end{eqnarray}
where $\sigma_{\mu\nu} = (i/2)[\gamma_\mu \gamma_\nu - \gamma_\nu \gamma_\mu]$.
Here $\Delta$ denotes a light--like vector, $\Delta.\Delta = 0$. $\psi$ denotes the quark field, 
$\lambda_r,~r \in [1, N_F^2-1]$ the $SU(N_F)$ matrices, $\gamma_\mu$ a Dirac matrix,
$D_\mu$ the covariant derivative and $A_\nu^a$ the 
gluon fields, with $F_{\mu\nu}^a$ the gluonic field strength tensor and $\ep_{\alpha \beta \gamma 
\delta}$ the Levi--Civita pseudo--tensor. For further notations we refer to Ref.~\cite{Blumlein:2021enk}.

Furthermore, the relative normalization between the quark--singlet and gluon external states has to be 
fixed. We adopt the same convention as in the massive on--shell case \cite{Bierenbaum:2009mv} by 
demanding 4--momentum conservation
\begin{eqnarray}
\int_0^1 dx x\left[\Sigma(x) + G(x)\right] = 1,
\end{eqnarray}
where $\Sigma$ is the quark singlet distribution
\begin{eqnarray}
\Sigma(x) = \sum_{k=1}^{N_F} \left[q_k(x) + \bar{q}_k(x)\right],
\end{eqnarray}
with $q$ the quark, $\bar{q}$ the antiquark distributions and $G$ is the gluon distribution.
In this way the operators are also fixed in the polarized case, replacing 
$\gamma_\mu$ by $\gamma_\mu \gamma_5$, etc., cf.~\cite{Blumlein:2021ryt}. 

We apply the following projectors to determine the different contributions to the OMEs. For external 
quark fields one uses in the unpolarized case
\begin{eqnarray}
\label{eq:Aiq}
\hat{A}_{iq} = \left[ \Delta \hspace*{-2.5mm} \slash \hat{A}_{iq}^{\rm phys} 
+ p \hspace*{-2mm} \slash \frac{\Delta.p}{p^2} 
\hat{A}_{iq}^{\rm eom} \right] (\Delta.p)^{N-1},~~~~i = q,g.
\end{eqnarray}
The quarkonic projections are obtained by
\begin{eqnarray}
\hat{A}_{iq}^{\rm phys} &=& \frac{1}{4 (\Delta.p)^{N}}~\text{tr}~\Biggl[\Biggl(p \hspace*{-2mm} \slash - 
\frac{p^2}{\Delta.p} 
\Delta 
\hspace*{-2.5mm} \slash\Biggr) \hat{A}_{iq} \Biggr]
\\
\hat{A}_{iq}^{\rm eom} &=& \frac{1}{(4 \Delta.p)^{N}}~\text{tr}~\Biggl[\Delta \hspace*{-2.5mm} 
\slash  \hat{A}_{iq} \Biggr].
\end{eqnarray}

For external gluons the decomposition is as follows \cite{Matiounine:1998ky}
\begin{eqnarray}
\hat{A}_{ig, \mu\nu} &=& 
  \hat{A}_{ig}^{\rm phys} T_{\mu\nu}^{(1)} 
+ \hat{A}_{ig}^{\rm eom}  T_{\mu\nu}^{(2)}
+ \hat{A}_{ig}^{\rm ngi}  T_{\mu\nu}^{(3)}, 
\end{eqnarray}
where
\begin{eqnarray}
T_{\mu\nu}^{(1)} &=& \frac{1}{2}[1+(-1)^N] \left[g_{\mu\nu} - \frac{p_\mu\Delta_\nu + \Delta_\mu 
p_\nu}{\Delta.p} 
+ \frac{\Delta_\mu \Delta_\nu p^2}{(\Delta.p)^2} \right] (\Delta.p)^N,
\\
T_{\mu\nu}^{(2)} &=& \frac{1}{2}[1+(-1)^N] \left[\frac{p_\mu p_\nu}{p^2} - \frac{p_\mu\Delta_\nu + 
\Delta_\mu p_\nu}{\Delta.p} 
+ \frac{\Delta_\mu \Delta_\nu p^2}{(\Delta.p)^2} \right] (\Delta.p)^N,
\\
T_{\mu\nu}^{(3)} &=& \frac{1}{2}[1+(-1)^N] \left[ - \frac{p_\mu\Delta_\nu + \Delta_\mu p_\nu}{2\Delta.p} 
+ \frac{\Delta_\mu \Delta_\nu p^2}{(\Delta.p)^2} \right] (\Delta.p)^N.
\end{eqnarray}
Later one more tensor structure is needed
\begin{eqnarray}
T_{\mu\nu}^{(4)} &=& \frac{1}{2}[1+(-1)^N] \left[ \frac{p_\mu\Delta_\nu + \Delta_\mu p_\nu}{2\Delta.p} 
\right] (\Delta.p)^N,
\end{eqnarray}
with \cite{Matiounine:1998ky}
\begin{equation}
\begin{array}{rcllrcl}
p^\mu T_{\mu\nu}^{(i)}       &=& 0,~~~ &(i=1,2),~~~   & p^\mu T_{\mu\nu}^{(i)}       &\neq& 0,~~~(i=3,4) 
\\ 
p^\mu p^\nu T_{\mu\nu}^{(i)} &=& 0,~~~ &(i=1,2,3),~~  & p^\mu p^\nu T_{\mu\nu}^{(4)} &\neq& 0.
\end{array}
\end{equation}
This tensor decomposition implies the choice of a physical gauge for the external gluon lines, with a 
propagator
\begin{equation}
D^{\mu\nu}(k^2) = i \frac{d^{\mu\nu}(k)}{k^2 + i 0},~~~d_{\mu\nu}(k) = -g^{\mu\nu} - n^2 \frac{k^\mu 
k^\nu}{(k.n)^2} + \frac{n^\mu k^\nu + n^\nu k^\mu}{k.n},
\end{equation}
and $n^2 \leq 0$. 

The resulting Ward identities are
\begin{eqnarray}
p^\mu \hat{A}_{qg, \mu\nu} &=& \frac{1}{2}[1+(-1)^N] \left[ -p_\nu + \frac{\Delta_\nu p^2}{\Delta.p}\right] 
(\Delta.p)^N 
\hat{A}_{qg}^{\rm ngi},
\\
p^\mu p^\nu \hat{A}_{qg, \mu\nu} &=& 0.
\end{eqnarray}
The gluonic projections are given by
\begin{eqnarray}
\hat{A}_{ig}^{\rm phys} &=& \frac{1}{D-2} \Biggl[g_{\mu\nu} + \frac{p^2}{(\Delta.p)^2} \Delta_\mu \Delta_\nu - 
\frac{p_\mu\Delta_\nu+p_\nu \Delta_\mu}{\Delta.p} 
\Biggr] \hat{A}_{ig}^{\mu\nu},
\\
\hat{A}_{ig}^{\rm eom} &=& \frac{p^2}{(\Delta.p)^2} \Delta_\mu \Delta_\nu \hat{A}_{ig}^{\mu\nu},
\\
\hat{A}_{ig}^{\rm ngi} &=& \Biggl[\frac{p_\mu p_\nu}{4 p^2}
- \frac{p_\mu\Delta_\nu+p_\nu \Delta_\mu}{2\Delta.p} 
\Biggr] \hat{A}_{ig}^{\mu\nu},
\\
\hat{A}_{ig}^{\rm wi} &=& \frac{p_\mu p_\nu}{4 p^2}\hat{A}_{ig}^{\mu\nu}.
\end{eqnarray}

In the polarized case the OMEs $\Delta \hat{A}_{iq}^{\rm l},~~i = q, g$ are given by
\begin{eqnarray}
\label{eq:Aiq}
\Delta \hat{A}_{iq} = \left[ \gamma_5 \Delta \hspace*{-2.5mm} \slash 
\Delta \hat{A}_{iq}^{\rm phys} + \gamma_5 p 
\hspace*{-2mm} 
\slash 
\frac{\Delta.p}{p^2} 
\Delta \hat{A}_{iq}^{\rm eom} \right] (\Delta.p)^{N-1},
\end{eqnarray}
and the OMEs with external gluon lines read
\begin{eqnarray}
\Delta \hat{A}_{ig, \mu\nu} &=& \ep_{\mu\nu\alpha\beta} \frac{\Delta^\alpha p^\beta}{\Delta.p} 
\Delta \hat{A}_{ig}^{\rm phys},~~~i = q,g.
\end{eqnarray}
To treat $\gamma_5$ in $D=4+\ep$ dimensions we use the Larin scheme \cite{Larin:1993tq,Matiounine:1998re}
and perform the replacement 
\begin{eqnarray}
  \slashed{p} \gamma_5 &=& \frac{i}{6} \epsilon_{\mu\nu\rho\sigma} p^\mu \gamma^\nu \gamma^\rho \gamma^\sigma ,
\end{eqnarray}
while contracting the occuring $\epsilon$--tensors in $D$ dimensions using
\begin{eqnarray}
  \epsilon_{\mu\nu\rho\sigma} \epsilon^{\alpha \lambda \tau \gamma} &=& - \text{Det}\left[ g_\omega^\beta \right],
  \qquad \beta = \alpha, \lambda, \tau, \gamma ; 
  \quad  \omega = \mu, \nu, \rho, \sigma ~.
\end{eqnarray}
In this scheme the projections onto the different terms are given by 
\begin{eqnarray}
  \Delta \hat{A}_{iq}^{\rm phys} &=& 
  - \frac{1}{4(D-2)(D-3)} \epsilon_{\mu\nu\rho\sigma} p^\rho \Delta^\sigma \text{tr}\left[ \slashed{p} \gamma^\mu \gamma^\nu \Delta \hat{A}_{iq} \right] (\Delta.p)^{-N-1}
  \nonumber \\ &&
  - \frac{p^2}{4(D-2)(D-3)} \epsilon_{\mu\nu\rho\sigma} p^\rho \Delta^\sigma \text{tr}\left[ \slashed{\Delta} \gamma^\mu \gamma^\nu \Delta \hat{A}_{iq} \right] (\Delta.p)^{-N-2}
  , \\ 
  \Delta \hat{A}_{iq}^{\rm eom} &=&
  \frac{p^2}{4(D-2)(D-3)} \epsilon_{\mu\nu\rho\sigma} p^\rho \Delta^\sigma \text{tr}\left[ \slashed{\Delta} \gamma^\mu \gamma^\nu \Delta \hat{A}_{iq} \right] (\Delta.p)^{-N-2}
  , \\
  \Delta \hat{A}_{ig}^{\rm phys} &=& \frac{1}{(D-2)(D-3)} \epsilon_{\mu\nu\rho\sigma} \Delta^\rho p^\sigma (\Delta.p)^{-N-1} \Delta \hat{A}_{ig}^{\mu\nu}
  .
\end{eqnarray}
There is no mixing between the physical and the ngi and the alien operators, because no symmetric 
rank two tensor contributes. 

{
For transversity the tensor decomposition for the OME reads 
\begin{eqnarray}
  \hat{A}_{qq,\mu}^{{\rm NS}, {\rm tr}} &=& \Delta^\rho \sigma_{\mu\rho} \tilde{A}_{qq}^{{\rm NS}, {\rm tr}}
  + c_1 \Delta_\mu + c_2 p_\mu + c_3 \gamma_\mu \slashed{p} + c_4 \Delta_\mu \slashed{\Delta} \slashed{p}
  + c_5 p_\mu \slashed{\Delta} \slashed{p}.
\end{eqnarray}
The physical OME can be extracted using the following projector
\begin{eqnarray}
  \tilde{A}_{qq}^{{\rm NS}, {\rm tr}} &=& \frac{i}{4(D-2)} \text{tr} \Biggl[
    \Biggl(
    - p^\mu \slashed{\Delta} \slashed{p}
    + \Delta.p \gamma^\mu \slashed{p}
    + i p^2 \Delta_\rho \sigma^{\mu \rho}
    \Biggr) \hat{A}_{qq,\mu}^{{\rm NS}, {\rm tr}} \Biggr] (\Delta.p)^{-N-2} 
    .
\end{eqnarray}

In the following we will insert the wave function renormalization \cite{WF} and  perform a partial 
renormalization of the coupling constant \cite{BETA} and the gauge parameter \cite{WF}, 
cf.~\cite{Blumlein:2021enk} for details. We will then denote the partly renormalized OMEs by 
$(\Delta) \tilde{A}_{ij}$.

The following representations hold for  $\tilde{A}_{ij}$ and $(\Delta) \tilde{A}_{ij}$. 
The different partly renormalized OMEs are given by
{
\begin{eqnarray}
\tilde{A}_{ij}^{\rm phys} 
&=& \delta_{ij} + \sum_{k = 1}^\infty S_\ep^k \left(\frac{-p^2}{\mu^2}\right)^{k \ep/2} {a}_s^k 
\tilde{A}_{ij}^{\rm phys,k} 
\\
\tilde{A}_{ij}^{\rm eom} 
&=& \sum_{k = 1}^\infty S_\ep^k \left(\frac{-p^2}{\mu^2}\right)^{k \ep/2} {a}_s^k 
\tilde{A}_{ij}^{\rm eom,k} 
\\
\tilde A_{ij}^{\rm ngi} 
&=&  \sum_{k = 1}^\infty S_\ep^k \left(\frac{-p^2}{\mu^2}\right)^{k \ep/2} {a}_s^k 
\tilde{A}_{ij}^{\rm ngi,k}, 
\end{eqnarray}
}
with $\mu$ the factorization scale and {the spherical factor $S_\ep$ is given by
\begin{eqnarray}
S_\ep =  \exp{\left[\frac{\ep}{2}\left( \gamma_E - \ln(4\pi) \right) \right]}  ~,
\end{eqnarray}
where $\gamma_E$ is the Euler-Mascheroni number.} Analogous relations hold in the polarized case and
the OMEs, cf.~\cite{Blumlein:2009rg,Blumlein:2021enk}, and those of transversity, which are
flavor non--singlet quantities. The structure of the partial amplitudes up to two--loop order is 
\begin{eqnarray}
\tilde{A}_{ij}^{\rm phys, 1}  &=& \frac{1}{\ep} \gamma_{ij}^{(0)} + a_{ij}^{(1,0)}
                            + \ep a_{ij}^{(1,1)} + \ep^2 a_{ij}^{(1,2)} + O(\ep^3),~~~~ij = qq~{\rm NS}, 
qg, gq, gg,
\\
\tilde{A}_{ij}^{\rm eom, 1}  &=& b_{ij}^{(1,0)}b+ \ep b_{ij}^{(1,1)} + \ep^2 b_{ij}^{(1,2)} + O(\ep^3)
~~~~ij = qq~{\rm NS}, qg, gq, gg,
\\
\tilde{A}_{ij}^{\rm ngi, 1}  &=& \frac{1}{\ep} \gamma_{gA}^{(0)} + c_{gA}^{(1,0)}
                            + \ep c_{gA}^{(1,1)} + \ep^2 c_{gA}^{(1,2)} + O(\ep^3),~~~~ij = gg,
\\
\tilde{A}_{ij}^{\rm phys, 2}  &=& 
  \frac{1}{\ep^2} a_{ij}^{(2,-2)} + \frac{1}{\ep} a_{ij}^{(2,-1)}
+ a_{ij}^{(2,0)} + \ep a_{ij}^{(2,1)} + O(\ep^2),~~~~ij = qq~{\rm NS}, qq~{\rm PS}, qg, gq, gg,
\nonumber\\
\\
\tilde{A}_{ij}^{\rm eom, 2}  &=& \frac{1}{\ep} b_{ij}^{(2,-1)}
+ b_{ij}^{(2,0)} + \ep b_{ij}^{(2,1)} + O(\ep^2),~~~~ij = qq~{\rm NS}, qq~{\rm PS}, qg, gq, gg,
\\
\tilde{A}_{ij}^{\rm ngi, 2}  &=& 
	  \frac{1}{\ep^2} c_{ij}^{(2,-2)} + \frac{1}{\ep} c_{ij}^{(2,-1)}
+ c_{ij}^{(2,0)} + \ep c_{ij}^{(2,1)} + O(\ep^2),~~~~ij = qg, gg.
\end{eqnarray}
Here the contributions $e^{(k,-1(-2))}$ with $e = a,b,c$ contain the LO and NLO anomalous dimensions, cf.
\cite{Matiounine:1998ky,Matiounine:1998re,Blumlein:2021ryt,Blumlein:2021enk}. Structures which lead to
new anomalous dimensions are dealt with in Section~\ref{sec:3}. In the polarized case there are no ngi 
contributions, but the eom terms for $ij = qq~{\rm NS, PS}$ and $gq$ to two--loop order. All results 
in the polarized case are presented in the Larin scheme \cite{Larin:1993tq,Matiounine:1998re}.

There are also 
eom contributions in the transversity cases, which we will not deal with in the present paper, since the tensor
composition in this case is even richer, cf.~\cite{Blumlein:2009rg}, but the anomalous dimensions come
only from the physical part, cf.~\cite{Blumlein:2021enk}. Note that our definitions differ in part from 
those in \cite{Matiounine:1998ky,Matiounine:1998re}. The corresponding mapping is obtained by performing the 
renormalization, which is carried out to 2--loop order in the present paper.
The anomalous dimensions obey the following expansion in the strong coupling constant
\begin{eqnarray}
\gamma_{ij}^{\rm a} = \sum_{k=1}^\infty {a}_s^k \gamma_{ij}^{(k-1),\rm a}.
\end{eqnarray}

Let us now turn to technical aspects of the calculation. The Feynman diagrams for the different operator 
insertions are generated by {\tt QGRAF} \cite{Nogueira:1991ex,Bierenbaum:2009mv}. The spinor and 
Lorentz--algebra is performed using {\tt FORM} \cite{FORM}. The different operator insertions are resummed 
using generating functions \cite{Ablinger:2014yaa} either for even or odd integer moments, 
cf.~\cite{Blumlein:2021enk}, implied by the respective crossing relations 
\cite{Politzer:1974fr,Blumlein:1996vs}. In this way the local operators reappear in terms of propagators and 
one may derive the integration--by--parts relations \cite{IBP} for the corresponding quantities, for which we use 
the package {\tt Crusher} \cite{CRUSHER}. For part of the calculation we performed the reduction using 
{\tt Litered} \cite{Lee:2012cn} and solved the differential equations by using the method of 
Refs.~\cite{Ablinger:2015tua,Ablinger:2018zwz}. The general solution followed the route described in 
Refs.~\cite{Blumlein:2021enk,Blumlein:2021ryt}. The relations between master integrals obtained by the reduction
using {\tt Crusher} allow to generated a sufficiently large number of Mellin moments by using the method of 
arbitrary high moments \cite{Blumlein:2017dxp} for the different color and zeta factors, which formed the 
basis for the method of guessing \cite{GUESS,Blumlein:2009tj}, implemented in {\tt Sage} \cite{SAGE,GSAGE}, to 
determine the corresponding recurrences. Those were solved by applying difference-ring theory \cite{Karr:81,
Schneider:01,Schneider:05a,Schneider:07d,Schneider:10b,Schneider:10c,Schneider:15a,Schneider:08c,CS:2021,EMSSP} 
as implemented in the package {\tt Sigma} \cite{SIG1,SIG2}. The generated recurrences in the present case 
factorize all at first order, unlike the case in a series of massive higher order calculations.
First one obtains solutions in terms of also cyclotomic harmonic sums \cite{Ablinger:2011te} because of 
separating even and odd moments. All the results can be written in terms of harmonic sums 
\cite{Vermaseren:1998uu,Blumlein:1998if} 
\begin{eqnarray} 
S_{b,\vec{a}}(N) = \sum_{k=1}^N \frac{({\rm sign}(b))^k}{k^{|b|}} S_{\vec{a}}(k),~~S_\emptyset = 1,~~a,b_i 
\in \mathbb{Z} \backslash \{0\},
\end{eqnarray}
for which one maps first 
\begin{eqnarray} 
S_{b,\vec{a}}(2N+1) = \frac{1}{2N+1} S_{\vec{a}}(2N+1) + S_{b,\vec{a}}(2N) 
\end{eqnarray} 
recursively and then applies the {\tt HarmonicSums} \cite{HARMSU, Blumlein:2003gb,
Ablinger:2011te, Ablinger:2013cf, Ablinger:2014bra,
Vermaseren:1998uu,
Blumlein:1998if,
Remiddi:1999ew,
Blumlein:2009ta} command {\tt Synchronize} and finally the command {\tt TransformToBasis[{\it expr}, Online 
$\rightarrow$ True]} is applied. All results are now obtained in terms of harmonic sums of argument $N$ only.

In the following section we turn now to the calculation of the matrix elements  of the so--called alien 
operators, which contribute  via mixing to the (ultraviolet) operator renormalization in the unpolarized case
and discusshh  in a subsequent section the mixing relations to two--loop order in explicit form.  
\section{The OMEs of the alien operators} 
\label{sec:3}

\vspace*{1mm} 
\noindent 
These operator matrix elements contribute in the unpolarized case and have been discussed in detail in 
Refs.~\cite{ Hamberg:1991qt, HAMBERG,Matiounine:1998ky}. The OMEs containing gluonic operators mix with 
these OMEs. The operators are given by
\begin{eqnarray}
O_A^{\mu_1,...,\mu_N}  &=& i^{N-2} {\rm\bf S}{\rm\bf Sp}\Biggl[
  F_{\alpha}^{a,\mu_1} D_{\alpha} \partial^{\mu_2} ... \partial^{\mu_{N-1}} A_{a}^{\mu_N}
  \nonumber \\ && \hspace{-0.5cm}
  + i g f^{abc}  F_{\alpha}^{a,\mu_1} \sum\limits_{i=2}^{N-1} \kappa_i \partial^\alpha
  \biggl\{
    \left( \partial^{\mu_2} ... \partial^{\mu_{i-2}} A_b^{\mu_{i-1}} \right)
    \left( \partial^{\mu_2} ... \partial^{\mu_{N-1}} A_c^{\mu_{N}} \right)
  \biggr\}
  \nonumber \\ && \hspace{-0.5cm}
  + {O}(g^3)
\Biggr] -~\text{trace~terms} ~,
\\
O_\omega^{\mu_1,...,\mu_N}  &=& i^{N-2} {\rm\bf S}{\rm\bf Sp}\Biggl[
  \xi^{a} \partial^{\mu_1} ... \partial^{\mu_{N-1}} \overline{\omega}^a
  \nonumber \\ && \hspace{-0.5cm}
  - i g f^{abc} \xi^{a} \sum\limits_{i=2}^{N-1} \eta_i \partial^{\mu_1}
  \biggl\{
    \left( \partial^{\mu_2} ... \partial^{\mu_{i-2}} \overline{\omega}_b \right)
    \left( \partial^{\mu_2} ... \partial^{\mu_{N-1}} A_c^{\mu_{N}} \right)
  \biggr\}
  \nonumber \\ && \hspace{-0.5cm}
  + {O}(g^3) 
\Biggr] -~\text{trace~terms} ~,
\\
O_B^{\mu_1,...,\mu_N} &=&  i^{N-1} {\rm\bf S}\Biggl[
  g \overline{\psi}_k \gamma_{\mu_1} (T_a)^{kl} A^{a,\mu_2} \partial^{\mu_3} ... \partial_{\mu_N} \psi_l
  + {O}(g^3)
\Biggr] -~\text{trace~terms} ,
\end{eqnarray}
with
\begin{eqnarray}
\kappa_i &=& \frac{(-1)^i}{8} + \frac{3}{8} \left[\frac{(n-2)!}{(i-1)! (n-i-1)!} - \frac{(n-2)!}{i! 
(N-i-2)!}
\right],
\\
\eta_i &=& \frac{(-1)^i}{4} + \frac{1}{4} \left[3\frac{(n-2)!}{(i-1)! (n-i-1)!} + \frac{(n-2)!}{i! 
(N-i-2)!}
\right]~,
\end{eqnarray}
and $\xi$, $\overline{\omega}$ denote the ghost and antighost respectively.
The Feynman rules for these operator insertions are summarized in Appendix~\ref{sec:A}.

To one loop order the OMEs are 
given by, after performing partial renormalization,
{
\begin{eqnarray}
\tilde{A}_{\rm Aq}^{\rm phys} &=& {a}_s 
S_\ep \left(\frac{-p^2}{\mu^2}\right)^{\ep/2} 
\Biggl[\frac{1}{\ep} \gamma_{\rm Aq}^{(0)} 
+ a_{\rm Aq}^{(1,0)}
+ \ep a_{\rm Aq}^{(1,1)}
+ \ep^2 a_{\rm Aq}^{(1,2)} \Biggr],
\\
\tilde{A}_{\rm Aq}^{\rm eom} &=& {a}_s S_\ep \left(\frac{-p^2}{\mu^2}\right)^{\ep/2} 
\Biggl[
b_{\rm Aq}^{(1,0)}
+ \ep b_{\rm Aq}^{(1,1)}
+ \ep^2 b_{\rm Aq}^{(1,2)} \Biggr],
\\
\tilde{A}_{\rm Bq}^{\rm phys} &=& {a}_s S_\ep \left(\frac{-p^2}{\mu^2}\right)^{\ep/2} 
\Biggl[
  -\frac{1}{\ep} \gamma_{\rm Aq}^{(0)} 
+ a_{\rm Bq}^{(1,0)}
+ \ep a_{\rm Bq}^{(1,1)}
+ \ep^2 a_{\rm Bq}^{(1,2)} \Biggr],
\\
\tilde{A}_{\rm Bq}^{\rm eom} &=& a_s S_\ep \left(\frac{-p^2}{\mu^2}\right)^{\ep/2} 
\Biggl[
b_{\rm Bq}^{(1,0)}
+ \ep b_{\rm Bq}^{(1,1)}
+ \ep^2 b_{\rm Bq}^{(1,2)} \Biggr],
\\
\tilde{A}_{\rm Ag}^{\rm phys} &=& {a}_s S_\ep \left(\frac{-p^2}{\mu^2}\right)^{\ep/2} 
\Biggl[
\frac{1}{\ep} \gamma_{\rm Ag}^{(0)} 
+ a_{\rm Ag}^{(1,0)}
+ \ep a_{\rm Ag}^{(1,1)}
+ \ep^2 a_{\rm Ag}^{(1,2)} \Biggr],
\\
\tilde{A}_{\rm Ag}^{\rm eom} &=& {a}_s S_\ep \left(\frac{-p^2}{\mu^2}\right)^{\ep/2} 
\Biggl[
b_{\rm Ag}^{(1,0)}
+ \ep b_{\rm Ag}^{(1,1)}
+ \ep^2 b_{\rm Ag}^{(1,2)} \Biggr],
\\
\tilde{A}_{\rm Ag}^{\rm ngi} &=& 1 + {a}_s S_\ep \left(\frac{-p^2}{\mu^2}\right)^{\ep/2} 
\Biggl[
\frac{1}{\ep} \gamma_{\rm AA}^{(0)} 
c_{\rm Ag}^{(1,0)}
+ \ep c_{\rm Ag}^{(1,1)}
+ \ep^2 c_{\rm Ag}^{(1,2)} \Biggr],
\\
\tilde{A}_{\rm Ag}^{\rm wi} &=& {a}_s S_\ep \left(\frac{-p^2}{\mu^2}\right)^{\ep/2} 
\Biggl[
d_{\rm Ag}^{(1,0)}
+ \ep d_{\rm Ag}^{(1,1)}
+ \ep^2 d_{\rm Ag}^{(1,2)} \Biggr],
\\
\tilde{A}_{\rm \omega g}^{\rm phys} &=&{a}_s S_\ep \left(\frac{-p^2}{\mu^2}\right)^{\ep/2} 
\Biggl[
- \frac{1}{\ep} \gamma_{\rm Ag}^{(0)} 
+ a_{\rm \omega g}^{(1,0)}
+ \ep a_{\rm \omega g}^{(1,1)}
+ \ep^2 a_{\rm \omega g}^{(1,2)} \Biggr],
\\
\tilde{A}_{\rm \omega g}^{\rm eom} &=&{a}_s S_\ep \left(\frac{-p^2}{\mu^2}\right)^{\ep/2} 
\Biggl[
+ b_{\rm \omega g}^{(1,0)}
+ \ep b_{\rm \omega g}^{(1,1)}
+ \ep^2 b_{\rm \omega g}^{(1,2)} \Biggr],
\\
\tilde{A} _{\rm \omega g}^{\rm ngi} &=& {a}_s S_\ep \left(\frac{-p^2}{\mu^2}\right)^{\ep/2} 
\Biggl[
  \frac{1}{\ep} \gamma_{\rm \omega A}^{(0)} 
+ c_{\rm \omega A}^{(1,0)}
+ \ep c_{\rm \omega A}^{(1,1)}
+ \ep^2 c_{\rm \omega A}^{(1,2)} 
\Biggr],
\\
\tilde{A}_{\rm \omega g}^{\rm wi} &=& {a}_s S_\ep \left(\frac{-p^2}{\mu^2}\right)^{\ep/2} 
\Biggl[ d_{\rm A g}^{(1,0)}
+ \ep d_{\rm A g}^{(1,1)}
+ \ep^2 d_{\rm A g}^{(1,2)} \Biggr],
\\
\tilde{A}_{\rm qg}^{\rm ngi} &=& {a}_s^2 S_\ep^2 \left(\frac{-p^2}{\mu^2}\right)^{\ep} 
\Biggl[ 
  \frac{1}{\ep^2} \gamma_{\rm qg}^{(0)} \gamma_{\rm gA}^{(0)} 
+ \frac{1}{\ep} \gamma_{\rm qg}^{(0)} a_{\rm gA}^{(1,0)} 
+ c_{qg}^{(2,0)}
+ \ep c_{qg}^{(2,1)} \Biggr],
\\
\tilde{A}_{\rm gg}^{\rm ngi} &=& {a}_s S_\ep \left(\frac{-p^2}{\mu^2}\right)^{\ep/2} 
\Biggl[ 
  \frac{1}{\ep} \gamma_{\rm gA}^{(0)} 
+ c_{gg}^{(1,0)}
+ \ep c_{gg}^{(1,1)} 
+ \ep^2 c_{gg}^{(1,2)} 
\Biggr].
\end{eqnarray}
}
Here we listed also those contributions to the OMEs given in Section~\ref{sec:2} for which new anomalous 
dimensions contribute to two--loop order. The latter read
{\begin{eqnarray}
        \gamma_{Aq}^{(0)} &=& - \frac{8 \textcolor{blue}{C_F}}{(N-1)N},
\\
        \gamma_{AA}^{(0)} &=& - \textcolor{blue}{C_A}
        \Biggl[
                \frac{16+46 N+N^2-12 N^3-3 N^4}{2 (N-1) N (1+N) (2+N)}
                + 2 \xi 
                + 6 S_1
        \Biggr],
\\
        \gamma_{Ag}^{(0)} &=& - \frac{2 \textcolor{blue}{C_A}}{(1+N) (2+N)},
\\
        \gamma_{\omega A}^{(0)} &=& -\frac{\textcolor{blue}{C_A} (N-1) (4+N) (6+N)}{6 N (1+N) (2+N)},
\\
        \gamma_{gA}^{(0)} &=&  \frac{4 \textcolor{blue}{C_A}}{(N-1)N}.
\end{eqnarray}
The higher order expansion terms are given by
}
\begin{eqnarray}
\label{eq:ALIEN1}
a_{Ag}^{(1,0)} &=&
\textcolor{blue}{C_A} \Biggl(
        \xi^2 \frac{1}{2 N}
        + \xi \frac{(1-2 N)}{(N-1) N}
        -\frac{Q_{9}}{(N-1) N (1+N)^2 (2+N)^2}
        +\frac{S_1}{(1+N) (2+N)}
\Biggr),
\\
a_{Ag}^{(1,1)} &=&
\textcolor{blue}{C_A} \Biggl(
        -\xi ^2 \Biggl[
                \frac{N+1}{4 N^2}
                +\frac{S_1}{4 N}
        \Biggr]
        +\xi  \Biggl[
                \frac{1-N+N^3}{2 (N-1)^2 N^2}
+\frac{(-1+2 N) S_1}{2 (N-1) N}
\Biggr]
\nonumber\\ &&  
        +\frac{S_1 Q_{9}}{2 (N-1) N (1+N)^2 (2+N)^2}               
        -\frac{Q_{17}}{(N-1)^2 N^2 (1+N)^3 (2+N)^3}
        -\frac{S_1^2}{4 (1+N) (2+N)}
\nonumber\\ &&  
        -\frac{3 S_2}{4 (1+N) (2+N)}
        +\frac{\zeta_2}{4 (1+N) (2+N)}
\Biggr),
\\
a_{Ag}^{(1,2)} &=&
\textcolor{blue}{C_A} \Biggl(
        \xi ^2 \Biggl[
                \frac{1+N}{8 N^3}
                +\frac{(1+N) S_1}{8 N^2}
                +\frac{S_1^2}{16 N}
                +\frac{3 S_2}{16 N}
                -\frac{\zeta_2}{16 N}
        \Biggr]
        +\xi  \Biggl[
                \frac{1-2 N+N^2-N^3}{4 (N-1)^3 N^3}
\nonumber\\ &&                
 +\frac{\big(
                        -1+N-N^3\big) S_1}{4 (N-1)^2 N^2}
                +\frac{(1-2 N) S_1^2}{8 (N-1) N}
                -\frac{3 (-1+2 N) S_2}{8 (N-1) N}
                +\frac{(-1+2 N) \zeta_2}{8 (N-1) N}
        \Biggr]
\nonumber\\ && 
        -\frac{S_1^2 Q_{9}}{8 (N-1) N (1+N)^2 (2+N)^2}
        -\frac{3 S_2 Q_{9}}{8 (N-1) N (1+N)^2 (2+N)^2}
\nonumber\\ &&         
+\frac{Q_{20}}{2 (N-1)^3 N^3 (1+N)^4 (2+N)^4}
        +\Biggl(
                \frac{Q_{17}}{2 (N-1)^2 N^2 (1+N)^3 (2+N)^3}
\nonumber\\ &&         
        +
                \frac{3 S_2}{8 (1+N) (2+N)}
        \Biggr) S_1
        +\frac{S_1^3}{24 (1+N) (2+N)}
        +\frac{7 S_3}{12 (1+N) (2+N)}
\nonumber\\ &&         
+\Biggl(
                \frac{Q_{9}}{8 (N-1) N (1+N)^2 (2+N)^2}
                -\frac{S_1}{8 (1+N) (2+N)}
        \Biggr) \zeta_2
\nonumber\\ &&         
-\frac{7 \zeta_3}{12 (1+N) (2+N)}
\Biggr),
\\
b_{Ag}^{(1,0)} &=& 
\textcolor{blue}{C_A} \Biggl(
        \xi  \Biggl[
                \frac{-1+9 N-6 N^2}{2 (N-1) N}
                +\frac{3}{2} S_1
        \Biggr]
        -\frac{2 \big(
                4+7 N+2 N^2\big)}{N (1+N) (2+N)}
        +\frac{(-2+N) \xi ^2}{4 N}
\Biggr),
\\
b_{Ag}^{(1,1)} &=&
\textcolor{blue}{C_A} \Biggl(
        \xi ^2 \Biggl[
                \frac{-2+2 N+6 N^2-5 N^3}{8 (N-1) N^2}
                +\frac{(1+N) S_1}{4 N}
        \Biggr]
+        \xi  \Biggl[
                \frac{Q_7}{4 (N-1)^2 N^2}
\nonumber\\ && 
                +\frac{(1-3 N) S_1}{4 (N-1) N}
                -\frac{3}{8} S_1^2
                -\frac{9}{8} S_2
        \Biggr]
        -\frac{Q_3}{N^2 (1+N)^2 (2+N)^2}
        +\frac{\big(
                4+7 N+2 N^2\big) S_1}{N (1+N) (2+N)}
\Biggr),
\\
b_{Ag}^{(1,2)} &=&
\textcolor{blue}{C_A} \Biggl(
        \xi ^2 \Biggl[
                \frac{Q_{10}}{16 (N-1)^2 N^3}
                +\frac{\big(
                        2-2 N-N^3\big) S_1}{16 (N-1) N^2}
                +\frac{(-1-N) S_1^2}{16 N}
                -\frac{3 (1+N) S_2}{16 N}
\nonumber\\ && 
                +\frac{(2-N) \zeta_2}{32 N}
        \Biggr]
        +\xi  \Biggl[
                \frac{Q_{11}
                }{8 (N-1)^3 N^3}
                +\Biggl(
                        \frac{1+8 N-13 N^2+6 N^3}{8 (N-1)^2 N^2}
                        +\frac{9}{16} S_2
                \Biggr) S_1
\nonumber\\ && 
                +\frac{(-1+3 N) S_1^2}{16 (N-1) N}
                +\frac{1}{16} S_1^3
                +\frac{3 (-1+3 N) S_2}{16 (N-1) N}
                +\frac{7}{8} S_3
                +\Biggl(
                        \frac{1-9 N+6 N^2}{16 (N-1) N}
                        -\frac{3}{16} S_1
                \Biggr) \zeta_2
        \Biggr]
\nonumber\\ &&         
+\frac{S_1 Q_3}{2 N^2 (1+N)^2 (2+N)^2}
        +\frac{Q_{13}}{N^3 (1+N)^3 (2+N)^3}
        +\frac{\big(
                -4-7 N-2 N^2\big) S_1^2}{4 N (1+N) (2+N)}
\nonumber\\ &&         
-\frac{3 \big(
                4+7 N+2 N^2\big) S_2}{4 N (1+N) (2+N)}
        +\frac{\big(
                4+7 N+2 N^2\big) \zeta_2}{4 N (1+N) (2+N)}
\Biggr),
\\
c_{Ag}^{(1,0)} &=&
\textcolor{blue}{C_A} \Biggl(
        \frac{\xi ^2}{2}
        +\xi  \Biggl[
                \frac{-2+2 N-N^2}{2 (N-1) N}
                -S_1
        \Biggr]
        +\frac{Q_{16}}{12 (N-1)^2 N^2 (1+N)^2 (2+N)^2}
\nonumber\\ && 
        +\frac{\big(
                8+20 N-N^2-3 N^3\big) S_1}{2 (N-1) N (1+N) (2+N)}
        +\frac{3}{2} S_1^2
        +\frac{9}{2} S_2
\Biggr),
\\
c_{Ag}^{(1,1)} &=&
\textcolor{blue}{C_A} \Biggl(
        -\frac{\xi ^2}{2}
        +\xi  \Biggl[
                \frac{-2+8 N-7 N^2+2 N^3}{4 (N-1)^2 N^2}
                +\frac{\big(
                        2-2 N+N^2\big) S_1}{4 (N-1) N}
                +\frac{1}{4} S_1^2
                +\frac{3}{4} S_2
                +\frac{1}{4} \zeta_2
        \Biggr]
\nonumber\\ && 
        +\frac{Q_{21}}{72 (N-1)^3 N^3 (1+N)^3 (2+N)^3}
        +\Biggl(
                \frac{Q_{15}}{8 (N-1)^2 N^2 (1+N)^2 (2+N)^2}
                -\frac{9}{4} S_2
        \Biggr) S_1
\nonumber\\ && 
        +\frac{\big(
                -8-20 N+N^2+3 N^3\big) S_1^2}{8 (N-1) N (1+N) (2+N)}
        -
        \frac{1}{4} S_1^3
        +\frac{3 \big(
                -8-20 N+N^2+3 N^3\big) S_2}{8 (N-1) N (1+N) (2+N)}
        -\frac{7}{2} S_3
\nonumber\\ &&         
+\Biggl(
                \frac{Q_1}{16 (N-1) N (1+N) (2+N)}
                +\frac{3}{4} S_1
        \Biggr) \zeta_2
\Biggr),
\\
c_{Ag}^{(1,2)} &=&
\textcolor{blue}{C_A} \Biggl(
        \xi ^2 \Biggl[
                \frac{1}{2}
                -\frac{\zeta_2}{16}
        \Biggr]
        +\xi  \Biggl[
                \frac{Q_2}{8 (N-1)^3 N^3}
                +\Biggl(
                        \frac{2-8 N+7 N^2-2 N^3}{8 (N-1)^2 N^2}
                        -\frac{3}{8} S_2
                \Biggr) S_1
                -\frac{7}{12} S_3
\nonumber\\ && 
                +\frac{\big(
                        -2+2 N-N^2\big) S_1^2}{16 (N-1) N}
                -\frac{3 \big(
                        2-2 N+N^2\big) S_2}{16 (N-1) N}
                +\Biggl(
                        \frac{2-2 N+N^2}{16 (N-1) N}
                        +\frac{S_1}{8}
                \Biggr) \zeta_2
                -\frac{1}{24} S_1^3
\nonumber\\ &&         
        -\frac{7}{12} \zeta_3
        \Biggr]
        -\frac{3 S_2 Q_{15}}{32 (N-1)^2 N^2 (1+N)^2 (2+N)^2}
        +\frac{Q_{23}}{432 (N-1)^4 N^4 (1+N)^4 (2+N)^4}
\nonumber\\ &&         
+\Biggl(
                \frac{Q_{19}}{16 (N-1)^3 N^3 (1+N)^3 (2+N)^3}
                -\frac{3 \big(
                        -8-20 N+N^2+3 N^3\big) S_2}{16 (N-1) N (1+N) (2+N)}
                +\frac{7}{4} S_3
        \Biggr) S_1
\nonumber\\ &&  
       +\Biggl(
                -\frac{Q_{15}}{32 (N-1)^2 N^2 (1+N)^2 (2+N)^2}
                +\frac{9}{16} S_2
        \Biggr) S_1^2
        +\frac{\big(
                8+20 N-N^2-3 N^3\big) S_1^3}{48 (N-1) N (1+N) (2+N)}
\nonumber\\ &&         
+\frac{1}{32}
         S_1^4
        +
        \frac{27}{32} S_2^2
        -\frac{7 \big(
                -8-20 N+N^2+3 N^3\big) S_3}{24 (N-1) N (1+N) (2+N)}
        +\frac{45}{16} S_4
\nonumber\\ &&         
+\Biggl(
                -\frac{Q_{16}}{96 (N-1)^2 N^2 (1+N)^2 (2+N)^2}
                +\frac{\big(
                        -8-20 N+N^2+3 N^3\big) S_1}{16 (N-1) N (1+N) (2+N)}
                -\frac{3}{16} S_1^2
\nonumber\\ &&         
        -\frac{9}{16} S_2
        \Biggr) \zeta_2
        +\Biggl(
                -\frac{7 Q_1}{48 (N-1) N (1+N) (2+N)}
                -\frac{7}{4} S_1
        \Biggr) \zeta_3
\Biggr),
\\
d_{Ag}^{(1,0)} &=& 
-\frac{\textcolor{blue}{C_A}}{4 N},
\\
d_{Ag}^{(1,1)} &=& 
\textcolor{blue}{C_A} \Biggl(
        \frac{1}{8 N^2}
        +\frac{S_1}{8 N}
\Biggr),
\\
d_{Ag}^{(1,2)} &=&
\textcolor{blue}{C_A} \Biggl(
        -\frac{1}{16 N^3}
        -\frac{S_1}{16 N^2}
        -\frac{S_1^2}{32 N}
        -\frac{3 S_2}{32 N}
        +\frac{\zeta_2}{32 N}
\Biggr),
\\
a_{Aq}^{(1,0)} &=&
\textcolor{blue}{C_F} \Biggl(
        -\frac{2 (-2+N) (-1+3 N)}{(N-1)^2 N^2}
        +\frac{(-4+N) \xi }{2 (N-1) N}
        +\frac{4 S_1}{(N-1) N}
\Biggr),
\\
a_{Aq}^{(1,1)} &=&
\textcolor{blue}{C_F} \Biggl(
        \xi  (N-4)  \Biggl[
                \frac{\big(1-3 N+N^2\big)}{4 (N-1)^2 N^2}
                -\frac{S_1}{4 (N-1) N}
        \Biggr]
        -\frac{(N-2) \big(
                -1+4 N-6 N^2+N^3\big)}{(N-1)^3 N^3}
\nonumber\\ &&  
       +\frac{(N-2) (-1+3 N) S_1}{(N-1)^2 N^2}
        -\frac{S_1^2}{(N-1) N}
        -\frac{3 S_2}{(N-1) N}
        +\frac{\zeta_2}{(N-1) N}
\Biggr),
\\
a_{Aq}^{(1,2)} &=&
\textcolor{blue}{C_F} \Biggl(
        \xi  (N-4)\Biggl[
                -\frac{ \big(
                        -1+4 N-6 N^2+2 N^3\big)}{8 (N-1)^3 N^3}
                -\frac{ \big(
                        1-3 N+N^2\big) S_1}{8 (N-1)^2 N^2}
                +\frac{ S_1^2}{16 (N-1) N}
\nonumber\\ && 
                +\frac{3  S_2}{16 (N-1) N}
                -\frac{\zeta_2}{16 (N-1) N}
        \Biggr]
+        \frac{(N-2) Q_4}{2 (N-1)^4 N^4}
        +\Biggl(
                \frac{3 S_2}{2 (N-1) N}
\nonumber\\ &&         
+        \frac{(N-2) \big(
                        -1+4 N-6 N^2+N^3\big)}{2 (N-1)^3 N^3}
        \Biggr) S_1
        -\frac{(N-2) (-1+3 N) S_1^2}{4 (N-1)^2 N^2}
        +\frac{S_1^3}{6 (N-1) N}
\nonumber\\ &&         
-\frac{3 (N-2) (-1+3 N) S_2}{4 (N-1)^2 N^2}
        +\frac{7 S_3}{3 (N-1) N}
        +\Biggl(
                \frac{(N-2) (-1+3 N)}{4 (N-1)^2 N^2}
                -\frac{S_1}{2 (N-1) N}
        \Biggr) 
\nonumber\\ &&  \times \zeta_2        
-\frac{7 \zeta_3}{3 (N-1) N}
\Biggr),
\\
b_{Aq}^{(1,0)} &=& 
-\xi \frac{\textcolor{blue}{C_F}}{N},
\\
b_{Aq}^{(1,1)} &=&
\xi \textcolor{blue}{C_F}  \Biggl(
        \frac{1-N}{2 N^2}
        +\frac{S_1}{2 N}
\Biggr),
\\
b_{Aq}^{(1,2)} &=&
\xi \textcolor{blue}{C_F}  \Biggl(
        \frac{N-1}{4 N^3}
        +\frac{(N-1) S_1}{4 N^2}
        -\frac{S_1^2}{8 N}
        -\frac{3 S_2}{8 N}
        +\frac{\zeta_2}{8 N}
\Biggr),
\\
a_{Bq}^{(1,0)} &=&
\textcolor{blue}{C_F} \Biggl(
         \xi \frac{2}{(N-1) N}
+        \frac{4 \big(
                1-3 N+N^2\big)}{(N-1)^2 N^2}
        -\frac{4 S_1}{(N-1) N}
\Biggr),
\\
a_{Bq}^{(1,1)} &=&
\textcolor{blue}{C_F} \Biggl(
        \xi  \Biggl[
                \frac{1-3 N+N^2}{(N-1)^2 N^2}
                -
                \frac{S_1}{(N-1) N}
        \Biggr]
        +\frac{2 \big(
                1-4 N+6 N^2-2 N^3\big)}{(N-1)^3 N^3}
\nonumber\\ &&    
     -\frac{2 \big(
                1-3 N+N^2\big) S_1}{(N-1)^2 N^2}
        +\frac{S_1^2}{(N-1) N}
        +\frac{3 S_2}{(N-1) N}
        -\frac{\zeta_2}{(N-1) N}
\Biggr),
\\
a_{Bq}^{(1,2)} &=&
\textcolor{blue}{C_F} \Biggl(
        \xi  \Biggl[
                \frac{1-4 N+6 N^2-2 N^3}{2 (N-1)^3 N^3}
                +\frac{\big(
                        -1+3 N-N^2\big) S_1}{2 (N-1)^2 N^2}
                +\frac{S_1^2}{4 (N-1) N}
                +\frac{3 S_2}{4 (N-1) N}
\nonumber\\ && 
                -\frac{\zeta_2}{4 (N-1) N}
        \Biggr]
        + \frac{Q_5}{(N-1)^4 N^4}
        +\Biggl(
                \frac{-1+4 N-6 N^2+2 N^3}{(N-1)^3 N^3}
                -\frac{3 S_2}{2 (N-1) N}
        \Biggr) S_1
\nonumber\\ &&         
+\frac{\big(
                1-3 N+N^2\big) S_1^2}{2 (N-1)^2 N^2}
        -\frac{S_1^3}{6 (N-1) N}
        +\frac{3 \big(
                1-3 N+N^2\big) S_2}{2 (N-1)^2 N^2}
        -\frac{7 S_3}{3 (N-1) N}
\nonumber\\ &&        
 +\Biggl(
                \frac{-1+3 N-N^2}{2 (N-1)^2 N^2}
                +\frac{S_1}{2 (N-1) N}
        \Biggr) \zeta_2
        +\frac{7 \zeta_3}{3 (N-1) N}
\Biggr),
\\
a_{\omega g}^{(1,0)} &=&
\textcolor{blue}{C_A} \Biggl(
        \frac{-4-N+N^2}{(1+N)^2 (2+N)^2}
        -\frac{S_1}{(1+N) (2+N)}
\Biggr),
\\
a_{\omega g}^{(1,1)} &=&
\textcolor{blue}{C_A} \Biggl(
        \frac{8+5 N-3 N^2-2 N^3}{(1+N)^3 (2+N)^3}
        +\frac{\big(
                4+N-N^2\big) S_1}{2 (1+N)^2 (2+N)^2}
        +\frac{S_1^2}{4 (1+N) (2+N)}
\nonumber\\ && 
        +\frac{3 S_2}{4 (1+N) (2+N)}
        -
        \frac{\zeta_2}{4 (1+N) (2+N)}
\Biggr),
\\
a_{\omega g}^{(1,2)} &=&
\textcolor{blue}{C_A} \Biggl(
        \frac{Q_6}{(1+N)^4 (2+N)^4}
        +\Biggl(
                \frac{-8-5 N+3 N^2+2 N^3}{2 (1+N)^3 (2+N)^3}
                -\frac{3 S_2}{8 (1+N) (2+N)}
        \Biggr) S_1
\nonumber\\ &&         
+\frac{\big(
                -4-N+N^2\big) S_1^2}{8 (1+N)^2 (2+N)^2}
        -\frac{S_1^3}{24 (1+N) (2+N)}
        +\frac{3 \big(
                -4-N+N^2\big) S_2}{8 (1+N)^2 (2+N)^2}
\nonumber\\ &&         
-\frac{7 S_3}{12 (1+N) (2+N)}
        +\Biggl(
                \frac{4+N-N^2}{8 (1+N)^2 (2+N)^2}
                +\frac{S_1}{8 (1+N) (2+N)}
        \Biggr) \zeta_2
\nonumber\\ &&         
+\frac{7 \zeta_3}{12 (1+N) (2+N)}
\Biggr),
\\
b_{\omega g}^{(1,0)} &=&
\frac{2 \textcolor{blue}{C_A}}{(1+N) (2+N)},
\\
b_{\omega g}^{(1,1)} &=&
\textcolor{blue}{C_A} \Biggl(
        \frac{-4-N+N^2}{(1+N)^2 (2+N)^2}
        -\frac{S_1}{(1+N) (2+N)}
\Biggr),
\\
b_{\omega g}^{(1,2)} &=&
\textcolor{blue}{C_A} \Biggl(
        \frac{8+5 N-3 N^2-2 N^3}{(1+N)^3 (2+N)^3}
        +\frac{\big(
                4+N-N^2\big) S_1}{2 (1+N)^2 (2+N)^2}
        +\frac{S_1^2}{4 (1+N) (2+N)}
\nonumber\\ && 
        +\frac{3 S_2}{4 (1+N) (2+N)}
        -\frac{\zeta_2}{4 (1+N) (2+N)}
\Biggr),
\\
c_{\omega g}^{(1,0)} &=&
\textcolor{blue}{C_A} \Biggl(
        -\frac{Q_{12}}{36 N^2 (1+N)^2 (2+N)^2}
        +\frac{\big(
                -4+2 N+N^2\big) S_1}{2 N (1+N) (2+N)}
\Biggr),
\\
c_{\omega g}^{(1,1)} &=&
\textcolor{blue}{C_A} \Biggl(
        -\frac{S_1 Q_{8}}{8 N^2 (1+N)^2 (2+N)^2}
        +\frac{Q_{18}}{216 N^3 (1+N)^3 (2+N)^3}
        +\frac{\big(
                4-2 N-N^2\big) S_1^2}{8 N (1+N) (2+N)}
\nonumber\\ && 
        -\frac{3 \big(
                -4+2 N+N^2\big) S_2}{8 N (1+N) (2+N)}
        +\frac{(N-1) (4+N) (6+N) \zeta_2}{48 N (1+N) (2+N)}
\Biggr),
\\
c_{\omega g}^{(1,2)} &=&
\textcolor{blue}{C_A} \Biggl(
        \frac{S_1^2 Q_{8}}{32 N^2 (1+N)^2 (2+N)^2}
        +\frac{3 S_2 Q_{8}}{32 N^2 (1+N)^2 (2+N)^2}
+\frac{Q_{22}}{1296 N^4 (1+N)^4 (2+N)^4}
\nonumber\\ &&         
        +\Biggl(
                \frac{Q_{14}}{16 N^3 (1+N)^3 (2+N)^3}
                +\frac{3 \big(
                        -4+2 N+N^2\big) S_2}{16 N (1+N) (2+N)}
        \Biggr) S_1
        +\frac{\big(
                -4+2 N+N^2\big) S_1^3}{48 N (1+N) (2+N)}
\nonumber\\ &&         
+\frac{7 \big(
                -4+2 N+N^2\big) S_3}{24 N (1+N) (2+N)}
        +\Biggl(
                \frac{Q_{12}}{288 N^2 (1+N)^2 (2+N)^2}
                +\frac{\big(
                        4-2 N-N^2\big) S_1}{16 N (1+N) (2+N)}
        \Biggr) \zeta_2
\nonumber\\ &&         
-\frac{7 (N-1) (4+N) (6+N) \zeta_3}{144 N (1+N) (2+N)}
\Biggr),
\\
d_{\omega g}^{(1,0)} &=&
\frac{\textcolor{blue}{C_A}}{4 N},
\\
d_{\omega g}^{(1,1)} &=&
\textcolor{blue}{C_A} \big(
        -\frac{1}{8 N^2}
        -\frac{S_1}{8 N}
\big),
\\
\label{eq:ALIEN2}
d_{\omega g}^{(1,2)} &=&
\textcolor{blue}{C_A} \Biggl(
        \frac{1}{16 N^3}
        +\frac{S_1}{16 N^2}
        +\frac{S_1^2}{32 N}
        +\frac{3 S_2}{32 N}
        -\frac{\zeta_2}{32 N}
\Biggr),
\end{eqnarray}


\noindent
with the polynomials
\begin{eqnarray}
\label{eq:POL3}
Q_1 &=& -3 N^4-12 N^3+N^2+46 N+16, 
\\  
Q_2 &=& -3 N^4+13 N^3-19 N^2+10 N-2, 
\\  
Q_3 &=& -N^4-13 N^3-30 N^2-24 N-8, 
\\  
Q_4 &=& 2 N^4-10 N^3+10 N^2-5 N+1, 
\\  
Q_5 &=& 3 N^4-10 N^3+10 N^2-5 N+1, 
\\  
Q_6 &=& 3 N^4+10 N^3+4 N^2-17 N-16, 
\\  
Q_7 &=& 12 N^4-30 N^3+25 N^2-8 N-1, 
\\  
Q_{8} &=& -3 N^5-14 N^4-11 N^3-24 N^2-60 N-16, 
\\  
Q_{9} &=& -2 N^5-13 N^4-40 N^3-53 N^2-28 N-8, 
\\  
Q_{10} &=& 12 N^5-26 N^4+17 N^3-6 N^2+4 N-2, 
\\  
Q_{11} &=& -24 N^6+72 N^5-64 N^4-3 N^3+29 N^2-7 N-1, 
\\  
Q_{12} &=& -8 N^6-21 N^5+22 N^4+3 N^3+184 N^2+540 N+144, 
\\  
Q_{13} &=& N^6-6 N^5-38 N^4-71 N^3-66 N^2-36 N-8, 
\\  
Q_{14} &=& -3 N^7-23 N^6-15 N^5+7 N^4-194 N^3-372 N^2-168 N-32, 
\\  
Q_{15} &=& -3 N^7+12 N^6+94 N^5+16 N^4-231 N^3-164 N^2-44 N+32, 
\\  
Q_{16} &=& -16 N^8-55 N^7-68 N^6-154 N^5+64 N^4+629 N^3+428 N^2+132 N-96, 
\\  
Q_{17} &=& -N^8-7 N^7+3 N^6+69 N^5+148 N^4+154 N^3+78 N^2-4 N-8, 
\\  
Q_{18} &=& -52 N^9-468 N^8-1635 N^7-2655 N^6-3027 N^5-2061 N^4+4822 N^3
\nonumber\\ &&
+10044 N^2
+4536 N+864, 
\\  
Q_{19} &=& 11 N^{10}+34 N^9-157 N^8-422 N^7+69 N^6+798 N^5+1025 N^4+670 N^3
\nonumber\\ &&
-308 N^2
-56 N+64, 
\\  
Q_{20} &=& -3 N^{11}-33 N^{10}-85 N^9-7 N^8+363 N^7+897 N^6+1085 N^5+527 N^4-96 N^3
\nonumber\\ &&
-88 N^2+16 N+16, 
\\  
Q_{21} &=& 92 N^{12}+552 N^{11}+729 N^{10}-1226 N^9-1623 N^8+3246 N^7+2599 N^6
\nonumber\\ &&
-5526 N^5
-10329 N^4-6766 N^3
+2772 N^2+504 N-576, 
\\  
Q_{22} &=& 320 N^{12}+3840 N^{11}+19840 N^{10}+57357 N^9+99480 N^8+114390 N^7
\nonumber\\ &&
+103660 N^6
+40845 N^5
-103420 N^4-176904 N^3-103680 N^2-34992 N
\nonumber\\ &&
-5184, 
\\  
\label{eq:POL4}
Q_{23} &=& -544 N^{16}-4352 N^{15}-10880 N^{14}+513 N^{13}+43712 N^{12}+41684 N^{11}
\nonumber\\ &&
-67480 N^{10}
-119386 N^9-6592 N^8+133644 N^7+221992 N^6+105361 N^5
\nonumber\\ &&
-75664 N^4-1512 N^3
+21600 N^2+1296 N-3456. 
\end{eqnarray}

The eom part of $A_{Bq}$ vanishes. In the above expressions the QCD color factor
are $\textcolor{blue}{C_F} = (N_c^2-1)/(2 N_c), \textcolor{blue}{C_A} 
= N_c, \textcolor{blue}{T_F} = 1/2$ for $SU(N_c)$ and $N_c = 3$ for QCD;
$\textcolor{blue}{N_F}$ denotes the number of massless quark flavors. $\zeta_k,~~k \in \mathbb{N},~k \geq 2$
are the values of Riemann's $\zeta$ function at integer arguments.
\section{The operator mixing in the unpolarized singlet case}
\label{sec:4}

\vspace*{1mm}
\noindent
In the calculation of the off--shell OMEs in the unpolarized singlet case mixing between the physical and 
alien operators occurs, which we discuss in the following. In Mellin $N$--space the $Z$--factor for the 
(ultraviolet) renormalization of the local operators up to ${O}(a_s^2)$ reads 
\begin{eqnarray}
  Z_{ij}^S &=& \delta_{ij} 
  + a_s S_\ep \frac{\gamma_{ij}^{(0)}}{\ep}
  + a_s^2 S_\ep^2 \Biggr[
    \frac{1}{\ep^2} 
    \left( 
      \frac{1}{2} \gamma_{il}^{(0)}\gamma_{lj}^{(0)} 
      + \beta_0 \gamma_{ij}^{(0)}  
    \right)
    + \frac{1}{2\ep} \gamma_{ij}^{(1)}
  \Biggr]
  + {O}(a_s^3)~.
\end{eqnarray}
The renormalized physical OMEs are obtained by\footnote{Note that we focus on the physical projection of 
the OMEs in this section, since the QCD anomalous dimensions are extracted from them. Similar relations 
can also be derived for the other projections.}
\begin{eqnarray}
  A_{ij}^{\rm phys} &=& (Z_{ik}^S)^{-1}  \tilde{\tilde{A}}_{kj},
\end{eqnarray}
with\footnote{The occurence of the factor of $-1/2$ in (\ref{eqmnhalfWVN}) is due to a convention used
in Ref.~\cite{Matiounine:1998ky}, {which is, however, not correctly implemented in 
Ref.~\cite{Matiounine:1998ky}.}}
\begin{eqnarray}
\tilde{\tilde{A}}_{qq}^{\rm PS} &=& {\tilde{A}}_{qq}^{\rm PS, phys},
\\
\tilde{\tilde{A}}_{qg} &=& {\tilde{A}}_{qg}^{\rm phys},
\\
\label{eq125}
\tilde{\tilde{A}}_{gq} &=& {\tilde{A}}_{gq}^{\rm phys} + \eta ( \tilde{A}_{Aq} + \tilde{A}_{B q}),
\\
\label{eqmnhalfWVN}
\tilde{\tilde{A}}_{gg} &=& {\tilde{A}}_{gg}^{\rm phys} 
- {\frac{\eta}{2}}( \tilde{A}_{Ag} + \tilde{A}_{\omega g}),
\end{eqnarray}
and
\begin{eqnarray}
  \eta = - a_s S_\ep \frac{\gamma_{gA}^{(0)}}{\ep} + {O}(a_s^2).
\end{eqnarray} 
The individual contributions $\tilde{\tilde{A}}_{ij}$ are given by
\begin{eqnarray}
\label{OMEsu1}
  \tilde{\tilde{A}}_{qq}^{{\rm PS}} &=& 
  a_s^2 S_\ep^2 \Biggl[
    \frac{1}{2\ep^2} \gamma_{gq}^{(0)} \gamma_{qg}^{(0)}
    + \frac{1}{\ep} 
    \Biggl( 
      \gamma_{qg}^{(0)} a_{gq}^{(1,0)}
      + \frac{\gamma_{qq}^{{\rm PS},(1)}}{2}
    \Biggr)
    + a_{qq}^{{\rm PS},(2,0)}
    + \ep \, a_{qq}^{{\rm PS},(2,1)}
  \Biggr] 
+ O(a_s^3)~, \\
  \tilde{\tilde{A}}_{qg} &=& 
  a_s S_\ep \Biggl[
    \frac{\gamma_{qg}^{(0)}}{\ep}
    + a_{qg}^{(1,0)}
    + \ep \, a_{qg}^{(1,1)}
    + \ep^2 \, a_{qg}^{(1,2)}
  \Biggr]
  + a_s^2 S_\ep^2 \Biggl[
    \frac{1}{2\ep^2} \gamma_{qg}^{(0)} 
    \Biggl( 
        \gamma_{gg}^{(0)} 
      + \gamma_{qq}^{(0)} 
      + 2 \beta_0 
    \Biggr)
    \nonumber \\ &&
    + \frac{1}{\ep}
    \Biggl(
      \gamma_{qg}^{(0)} a_{gg}^{(1,0)}
      + \gamma_{qq}^{(0)} a_{qg}^{(1,0)}
      + \frac{\gamma_{qg}^{(1)}}{2}
    \Biggr)
    + a_{qg}^{(2,0)}
    + \ep \, a_{qg}^{(2,1)}
  \Biggr]
+ O(a_s^3)~, \\
  \tilde{\tilde{A}}_{gq} &=& 
  a_s S_\ep \Biggl[
    \frac{\gamma_{gq}^{(0)}}{\ep} 
    + a_{gq}^{(1,0)}
    + \ep \, a_{gq}^{(1,1)}
    + \ep^2 \, a_{gq}^{(1,2)}
  \Biggr]
  + a_s^2 S_\ep^2 \Biggl[
    \frac{1}{2\ep^2} \gamma_{gq}^{(0)} 
    \Biggl( 
      \gamma_{gg}^{(0)} + \gamma_{qq}^{(0)} + 2 \beta_0
    \Biggr)
    \nonumber \\ &&
    + \frac{1}{\ep}
    \Biggl( 
      \gamma_{gg}^{(0)} a_{gq}^{(1,0)}
      + \gamma_{gq}^{(0)} a_{qq}^{(1,0)}
      + \frac{\gamma_{gq}^{(1)}}{2}\Biggr)
      + \frac{\delta_{gq}^{(-1)}}{\ep}
    + a_{gq}^{(2,0)}
    + \delta_{gq}^{(0)}
    + \ep \, a_{gq}^{(2,1)}
    + \ep \delta_{gq}^{(1)}
  \Biggr]
+ O(a_s^3)~, 
\nonumber\\
\\
\label{OMEsu4}
  \tilde{\tilde{A}}_{gg} &=&
  1 + a_s S_\ep \Biggl[ 
    \frac{\gamma_{gg}^{(0)}}{\ep}
    + a_{gg}^{(1,0)}
    + \ep \, a_{gg}^{(1,1)}
    + \ep^2 \, a_{gg}^{(1,2)}
  \Biggr]
  + a_s^2 S_\ep^2 \Biggl[
    \frac{1}{2\ep^2} \Biggl( 
      \gamma_{gg}^{(0)} \left( \gamma_{gg}^{(0)} + 2 \beta_0 \right)
      \nonumber\\ &&  
      + \gamma_{gq}^{(0)} \gamma_{qg}^{(0)}
    \Biggr)
    + \frac{1}{\ep}
    \Biggl(
      \gamma_{gg}^{(0)} a_{gg}^{(0)}
      + \gamma_{gq}^{(0)} a_{qg}^{(1,0)}
      + \frac{\gamma_{gg}^{(1)}}{2} \Biggr)
      + \frac{\delta_{gg}^{(-1)}}{\ep}
    + a_{gg}^{(2,0)}
    + \delta_{gg}^{(0)}
\nonumber\\ &&
    + \ep \, a_{gg}^{(2,1)}
    + \ep \delta_{gg}^{(1)}
  \Biggr]
+ O(a_s^3)~.
\end{eqnarray}
In the flavor non--singlet cases, including transversity \cite{Blumlein:2021enk}, and in the polarized 
singlet case \cite{Blumlein:2021ryt} no mixing with the alien operators occurs. To two--loop order, these
terms contribute in the renormalization of $\tilde{A}_{gq}$ and $\tilde{A}_{gg}$, since one has to consider 
the combinations (\ref{eq125}, \ref{eqmnhalfWVN}) in this case.
Because of this the additional contributions 
{
\begin{eqnarray}
  \delta_{gq}^{(i-1)} &=& - \gamma_{gA}^{(0)} \left( a_{Aq}^{(1,i)} + a_{Bq}^{(1,i)} \right),~~~i \geq 0 
~,
\\
  \delta_{gg}^{(i-1)} &=&  \frac{\gamma_{gA}^{(0)}}{2} \left( a_{Ag}^{(1,i)} + a_{\omega g}^{(1,i)} \right),~~~i 
\geq 0
\end{eqnarray}
}
are present. The anomalous dimensions are obtained from the pole terms of (\ref{OMEsu1}--\ref{OMEsu4}). 
\section{The unpolarized OMEs} 
\label{sec:5}

\vspace*{1mm} 
\noindent 
We now turn to the calculation of contributions to the unpolarized off--shell OMEs for non--negative
order in the dimensional parameter $\ep$. They are gauge dependent in general and are given 
by\footnote{The polynomials are listed in Appendix~\ref{sec:B}.}



\section{The polarized OMEs} 
\label{sec:6}

\vspace*{1mm} 
\noindent 
The expansion coefficients in the polarized flavor singlet case are calculated using the Larin scheme 
\cite{Larin:1993tq,Matiounine:1998re}, 
as the complete calculation of the massless off--shell OMEs in the polarized case.
By this one obtains the anomalous dimensions in the Larin scheme, which are finally transformed
into the {\sf M} scheme \cite{Matiounine:1998re,Moch:2014sna}.
{
  In the non--singlet case one could use a known Ward identity and obtain the anomalous dimension 
  directly in the $\sf \overline{MS}$ scheme. However, the consitent renormalization of the 
  singlet case requires to calculate the non-singlet also in the Larin scheme.
}


\section{The Transversity OMEs} 
\label{sec:7}

\vspace*{1mm} 
\noindent
In the case of transversity various OMEs contribute \cite{Blumlein:2009rg,Blumlein:2021enk}. 
Here we only consider the expansion coefficients contributing to the physical projection
$A_{qq}^{\rm tr, \pm}$. The coefficients corresponding to the non--negative powers in $\ep$ are given by



In summary, the expressions shown in Sections~\ref{sec:3}--\ref{sec:6} depend on the following 
28 harmonic sums after algebraic reduction \cite{Blumlein:2003gb},
\begin{eqnarray}
&&\big\{S_{-5},S_{-4},S_{-3},S_{-2},S_1,S_2,S_3,S_4,S_5,S_{-4,1},S_{-3,1},S_{-2,1},S_{-2,2},S_{-2,3},
S_{2,-3},S_{2,1},S_{2,3},S_{3,1},
\nonumber\\ &&
S_{4,1},
S_{-3,1,1},S_{-2,1,1},S_{-2,2,1},S_{2,1,-2},S_{2,1,1},
S_{2,2,1},S_{3,1,1},S_{-2,1,1,1},S_{2,1,1,1}\big\}.
\end{eqnarray}
After applying also the structural relations \cite{Blumlein:2009ta} the following 13 sum contribute,
\begin{eqnarray}
\{S_1, 
S_{-2, 1}, 
S_{ 2, 1}, 
S_{-3, 1}, 
S_{ 4, 1}, 
S_{-4, 1}, 
S_{-2, 1, 1}, 
S_{ 2, 1, 1}, 
S_{-3, 1, 1}, 
S_{-2, 2, 1}, 
S_{ 2, 2, 1}, 
S_{-2, 1, 1, 1}, 
S_{ 2, 1, 1, 1}\}.
\end{eqnarray}
\section{Comparison to the literature} 
\label{sec:8}

\vspace*{1mm} 
\noindent 
In Refs.~\cite{Matiounine:1998ky,Matiounine:1998re}, extending erarlier results in Refs.~\cite{Hamberg:1991qt,
HAMBERG}, unpolarized and polarized OMEs have been calculated
to $O(a_s^2 \ep^0)$, i.e. one order in $\ep$ less than in the present calculation and the OMEs of transversity 
were not considered. The calculation has been carried out in $z$--space. This has been slightly before
harmonic sums \cite{Vermaseren:1998uu,Blumlein:1998if} and harmonic polylogarithms \cite{Remiddi:1999ew} 
became the standard entities to represent different calculation steps and the final results in single scale 
calculations. Therefore classical polylogarithms and Nielsen integrals 
\cite{NIELSEN,KMR70,Kolbig:1983qt,Devoto:1983tc,LEWIN1,LEWIN2}, partly with involved argument, were used. 
In \cite{Matiounine:1998ky,Matiounine:1998re} the expansion coefficients of the completely 
unrenormalized OMEs are discussed, while we perform the renormalization of the strong coupling and the gauge 
parameter first. Furthermore, the gauge parameter there is $\tilde{\xi} = 1 - \xi$, compared to the present 
case. In the following we summarize a series of differences to the present results which we have
observed, both due to typographical and combinatoric errors. We will mention main observations only and 
do not intend to give a complete list.

Comparing to \cite{Matiounine:1998ky,Matiounine:1998re} one first observes that the concrete results
for the non--singlet OMEs differ by a global minus sign.
For all the other OMEs we find a relative minus sign at ${O}(\alpha_s)$, but agree at 
${O}(\alpha_s^2)$. 
The formal representations in terms of anomalous dimensions,
i.e. Eqs.~(2.8, 2.16, 2.31, 2.34) of Ref.~\cite{Matiounine:1998ky}
and Eqs.~(2.18, 2.24, 2.26, 2.29) of Ref.~\cite{Matiounine:1998re}, are, however, correct.
Furthermore, 
the expressions for $\Delta \hat{A}_{iq}^{\rm phys}$ there are obtained as the sum $\Delta 
\hat{A}_{iq}^{\rm phys} + \Delta \hat{A}_{iq}^{\rm eom}$ of the present results in the Larin scheme. 

After adjusting the signs we find also some specific differences at the one--loop level.
Let us define
\begin{eqnarray}
\delta A_{ij}^{(k)}(x) &=& A_{ij}^{(k),\rm this~calc.}(x) - A_{ij}^{(k),\rm [19]}(x).
\\
\delta \Delta A_{ij}^{(k)}(x) &=& \Delta A_{ij}^{(k),\rm this~calc.}(x) - \Delta A_{ij}^{(k),\rm 
[20]}(x).
\end{eqnarray}
We obtain the following differences
\begin{eqnarray}
\label{err1}
\delta A_{gg}^{(1)}(z) &=&  
-20 \textcolor{blue}{C_A} \frac{\HA_0 - \HA_1}{1-z} {+ 20 \delta(1-z)}
-\frac{1}{12} \textcolor{blue}{C_A} \ep (3 \xi \zeta_2 + 8 \zeta_3) \delta(1-z)
\\
\delta \Delta A_{gg}^{(1)}(z) &=&  
-\frac{1}{12} \textcolor{blue}{C_A} \ep (3 \xi \zeta_2 + 8 \zeta_3) \delta(1-z),
\end{eqnarray}
where the harmonic polylogarithms \cite{Remiddi:1999ew} are defined by $\HA_{\vec{a}} \equiv \HA_{\vec{a}}(z)$.
Note that the $O(\ep^0)$ contribution in (\ref{err1}) is not in agreement with the earlier calculation 
\cite{Hamberg:1991qt,HAMBERG}, where the corresponding expansion has been given to $O(\ep^0)$. It is 
therefore rather difficult to use these older results in current calculations.
\section{Conclusions} 
\label{sec:9}

\vspace*{1mm} 
\noindent 
We have calculated the unpolarized and polarized massless off--shell twist--2 operator matrix elements
in QCD to two--loop orders in an automated way. These quantities are of interest because they allow the direct
calculation of the QCD anomalous dimensions. Contrary to this, on shell massive operator matrix elements only
allow to compute the anomalous dimensions of contributions $\propto T_F$ in the same order in perturbation 
theory and require the corrections of one  more order in the coupling constant to obtain the complete anomalous 
dimensions. The off--shellness implies gauge variant expressions in general. In particular, the equation of motion 
is not valid anymore for these terms and non--gauge invariant contributions are present, which may mix with the 
gauge invariant contributions. Despite of these technical complications, the method allows for a direct calculations 
of the anomalous dimensions. The on--shell calculation of the forward Compton amplitude also allows to compute 
the anomalous dimensions contributing to the deep--inelastic structure functions. However, special arrangements 
are necessary for the gluonic contributions, which require auxiliary Higgs- and graviton fields, also 
complicating this method.

We compared our results to those in Refs.~\cite{Matiounine:1998ky,Matiounine:1998re} for the contributions to 
$O(a_s^2 \ep^0)$ and calculated newly the contributions of  $O(a_s \ep^2)$ and $O(a_s^2 \ep)$, which contribute 
to the calculation of the four--loop anomalous dimensions. In the comparison we have found a series of typographical
and combinatoric errors in the previous calculations \cite{Matiounine:1998ky,Matiounine:1998re}. In particular, 
as a by--product of the present calculation all unpolarized and the polarized anomalous dimensions and those for 
transversity are correctly obtained to two--loop order \cite{Floratos:1977au,
GonzalezArroyo:1979he,
GonzalezArroyo:1979ng,
GonzalezArroyo:1979df,
Curci:1980uw,
Furmanski:1980cm,
Floratos:1981hs,
Hamberg:1991qt,
Mertig:1995ny,
SP_PS1,
Ellis:1996nn,
Ablinger:2014lka,
Ablinger:2014vwa,
Ablinger:2014nga,
Ablinger:2017tan,
Behring:2019tus,
Blumlein:2021enk,
Blumlein:2021ryt}.
Our calculation has been performed in an algorithmic manner using well established methods having been applied 
in various massless and massive three--loop calculations before, as e.g. \cite{Ablinger:2017tan,Behring:2019tus,
Ablinger:2017ptf}. The present formalism allows extensions to higher loop calculations. Here, however, also new 
structures are expected to emerge and the complexity of the calculation will naturally be larger.
The quantities to be computed are expressible in terms of the usual harmonic sums \cite{Vermaseren:1998uu,
Blumlein:1998if} and no other function spaces as e.g.~\cite{Ablinger:2011te,Ablinger:2013cf,Ablinger:2014bra,
Ablinger:2017bjx} are required to derive and express the results such as the different expansion coefficients 
of the different massless OMEs and the anomalous dimensions. 

\vspace*{5mm} 
\noindent 
{\bf Acknowledgments.}\\ 
We would like to thank A.~De Freitas and G.~Sborlini for discussions. This project has received 
funding from the European Union's Horizon 2020 research and innovation programme under the Marie 
Sk\/{l}odowska-Curie grant agreement No. 764850, SAGEX, and from the Austrian Science Fund (FWF)
grant SFB F50 (F5009-N15) and P33530.

\appendix
\section{Feynman rules 
\label{sec:A}}

\vspace*{1mm}
\noindent
The alien operators introduce new Feynman rules for operator insertions.
They have been given in Refs.~\cite{Hamberg:1991qt,HAMBERG,Matiounine:1998ky}.
We give them here fore completeness and correct typographical errors; particularly all sums 
have to start with the lower index 0. 
\begin{eqnarray}
        O_{A,\mu\nu}^{ab}(p,-p) &=& \delta^{ab} \frac{1+(-1)^N}{2}
        \left[ 
                2 \Delta_\mu \Delta_\nu p^2 
                - ( p_\mu \Delta_\nu + p_\nu \Delta_\mu ) \Delta.p 
        \right] (\Delta.p)^{N-2} ,
        \\ 
        O_{A,\mu\nu\lambda}^{abc}(p,q,k) &=& -ig \frac{1+(-1)^N}{2} f^{abc} V_{\mu\nu\lambda}^{(3)}(p,q,k),
\end{eqnarray}

\begin{eqnarray}
        O_{\omega}^{ab}(p,-p) &=& - \delta^{ab} (\Delta.p)^{N} ,
        \\ 
        O_{\omega,\mu}^{abc}(p,q,k) &=& ig f^{abc} \tilde{V}_{\mu}^{(3)}(p,q,k),
\end{eqnarray}

\begin{eqnarray}
        V_{\mu\nu\lambda}^{(3)}(p,q,k) &=& 
        \left[  
                  \Delta_\mu \Delta_\nu (q_\lambda-p_\lambda)
                + \Delta_\mu \Delta_\lambda (p_\nu-k_\nu)
                + g_{\nu\lambda} \Delta_\mu ( \Delta.k - \Delta.q )
        \right] (\Delta.p)^{N-2}
        \nonumber \\ &&
        - \frac{1}{4} \Delta_\nu \Delta_\lambda ( \Delta_\mu p^2 - p_\mu \Delta.p )
        \sum\limits_{i=0}^{N-3}
        \biggl[
                (-\Delta.q)^i (\Delta.k)^{N-3-i}
                - 3 (-\Delta.k)^i (\Delta.p)^{N-3-i}
        \nonumber \\ &&
                - 3 (-\Delta.p)^i (\Delta.q)^{N-3-i}
        \biggr]
        + 
        \begin{Bmatrix}
                p \rightarrow q \rightarrow k \rightarrow p \\
                \mu \rightarrow \nu \rightarrow \lambda \rightarrow \mu
        \end{Bmatrix}
        + 
        \begin{Bmatrix}
                p \rightarrow k \rightarrow q \rightarrow p \\
                \mu \rightarrow \lambda \rightarrow \nu \rightarrow \mu
        \end{Bmatrix}
        \\
        \tilde{V}_{\mu}^{(3)}(p,q,k) &=&
        \Delta_\mu 
        \Biggl[
                \frac{1}{2} (\Delta.k)^{N-1}
                + \frac{1}{4} (\Delta.q - 3 \Delta.k) (\Delta.q)^{N-2}
                + \frac{1}{4} (\Delta.k - \Delta.q) (\Delta.p)^{N-2}
        \nonumber \\ &&
                - \frac{3}{4} (\Delta.k)^{2} \sum\limits_{i=0}^{N-3} (\Delta.k)^i (-\Delta.q)^{N-3-i}
        \Biggr].
\end{eqnarray}
\section{The polynomials 
\label{sec:B}}

\vspace*{1mm}
\noindent
In the following we list the polynomials occurring in Eqs.~(\ref{eq:UNP1}--\ref{eq:UNP2}, 
\ref{eq:POL1}--\ref{eq:POL2}) and (\ref{eq:TR1}--\ref{eq:TR2}).




\end{document}